\documentclass[twocolumn, showpacs, secnumarabic, amssymb, aps, prd]{revtex4}

\newcommand{\bi}{\bibitem}
\newcommand{\be}{\begin{eqnarray}}
\newcommand{\ee}{\end{eqnarray}}
\newcommand{\rar}{\rightarrow}

\begin{document}

\title{Dark Energy and the mass of galaxy clusters}

\author{Cosimo Bambi}
\email[E-mail: ]{bambi@fe.infn.it}
\affiliation{Istituto Nazionale di Fisica Nucleare, 
Sezione di Ferrara, I-44100 Ferrara, Italy\\
Dipartimento di Fisica, Universit\`a degli Studi di Ferrara,
I-44100 Ferrara, Italy}

\date{\today}

\begin{abstract}
Up to now, Dark Energy evidences are based on the
dynamics of the universe on very large scales,
above 1 Gpc. Assuming it continues to behave like
a cosmological constant $\Lambda$ on much smaller scales,
I discuss its effects on the motion of non-relativistic 
test-particles in a weak gravitational field and I propose 
a way to detect evidences of $\Lambda \neq 0$ at the 
scale of about 1 Mpc: the main ingredient is the measurement 
of galaxy cluster masses.
\end{abstract}


\maketitle

\section{Introduction}

Present observational data \cite{data} suggest
that about 70\% of the energy in the universe is made
of a mysterious substance, the so called Dark Energy,
which would be unable to form structures and whose 
energy density would be constant in space and time \cite{carroll}.
However, these conclusions are essentially based on the 
dynamics of the universe on very large scales, above 1 Gpc.

In the simplest case, Dark Energy would be the cosmological
constant, maybe somehow related to the vacuum energy, and 
hence uniform on macroscopic distances. According to other
proposals, it could instead be the energy of some new
weakly interacting field, so it may not have an exactly
homogeneous and isotropic distribution and it may
be capable of clustering, even if not like standard matter. 
It would be therefore very important to observe Dark 
Energy effects on smaller scales, in order to distinguish
different pictures and reject unsuccessful Dark Energy
candidates. This becomes even more relevant by the light
of the possibility that Dark Energy does not exist and
that the accelerated expansion rate of the universe can
be explained with standard physics \cite{riotto} or 
with modifications of General Relativity \cite{mgr}.

Effects of a non-null cosmological constant $\Lambda$ on
the Solar System have been considered and upper bounds on 
its local value have been deduced \cite{ssystem, sereno}, 
but they are far from the value we get from cosmological 
observations.

In this paper I discuss the effects of a non zero
cosmological constant in the ``Newtonian limit'',
i.e. under the approximations of slow motion and weak
gravitational field. In particular, I focus the attention
on the dynamics of systems such as galaxy clusters, which 
could be used to probe distances of about 1 Mpc, and I
show the possibility of observing Dark Energy evidences
by precise measurements of their gravitational masses.
The existence of a cosmological constant at this 
relatively small scale could also be checked observing the
``local expansion rate'' of the universe (as opposed to
the ``global expansion rate''), as suggested in 
Ref. \cite{chernin}.

The content of the paper is as follows. In the next
section I derive the effective gravitational force acting on
a non-relativistic test-particle in the case of a non zero
cosmological constant and in Section \ref{s-discussion} I 
discuss general features and implications. 
In Section \ref{s-implications} I consider cosmological
constant effects on the measurements of galaxy cluster 
masses and I show the possibility of getting evidences of 
$\Lambda$ at the scale of about 1 Mpc. 
In Section \ref{s-conclusion} there are final remarks 
and conclusion.

\section{Newtonian limit}
\label{s-isl}

The Kottler spacetime \cite{kottler}, also known 
as Schwarzschild-de Sitter spacetime, is the
unique spherically symmetric solution of Einstein's
vacuum field equation with a cosmological constant 
$\Lambda$. In static and spherically symmetric 
coordinates, the line element is \cite{book}
\be\label{metric}
ds^2 = A(r)dt^2 -\frac{dr^2}{A(r)} 
-r^2\left(d\theta^2 + \sin^2\theta d\phi^2\right) \, ,
\ee
where
\be
A(r) = 1 - \frac{2G_NM}{r} - \frac{\Lambda}{3}r^2 \,
\ee
and $M$ is the mass of the source. Eq. (\ref{metric}) 
reduces to the standard Schwarzschild metric for 
$\Lambda=0$ and to the de Sitter one for $M=0$; in the 
latter case the coordinates are not the ones often used in
cosmology, where the expansion of the spacetime is explicit. 

The gravitational force acting on a test-particle
in the Newtonian limit can be deduced as follows 
(see e.g. Ref. \cite{weinberg}). First, we write the 
geodesic equation, which describes the motion of a 
test-particle in a background gravitational field
\be\label{geodesic}
\frac{d^2x^\sigma}{d\lambda^2} 
+ \Gamma^{\sigma}_{\mu\nu}
\frac{dx^\mu}{d\lambda}\frac{dx^\nu}{d\lambda} = 0 \, .
\ee
Here and in the following, Greek letters $\mu, \,\nu, \,...$
($\mu=0,\,1,\,2,\,3$) denote spacetime indices, Latin 
letters $i, \,j, \,...$ ($i=1,\,2,\,3$) denote
space indices and $\lambda$ is an affine parameter.
In the ``Newtonian limit'' we assume that
the motion of the test-particle is slow and that
the gravitational field is weak. The first hypothesis
means
\be
\frac{dt}{d\lambda} \gg \frac{dx^i}{d\lambda} \, ,
\ee
while the assumption of weak gravitational field lets us write 
the metric tensor as $g_{\mu\nu} = \eta_{\mu\nu} + h_{\mu\nu}$,
where $\eta_{\mu\nu}$ is the Minkowski metric and
$h_{\mu\nu}$ a small perturbation. To first order in 
$dx^i/d\lambda$ and $h_{\mu\nu}$ and for time independent
metrics, like the one in Eq. (\ref{metric}), the geodesic 
equation can be written as
\be\label{nlimit}
\frac{d^2{\bf r}}{dt^2} = -\frac{1}{2} \nabla h_{tt} \, ,
\ee
where ${\bf r}$ is the flat position 3-vector of the test-particle
and $\nabla$ is the flat nabla operator. In the case of the 
Kottler metric, Eq. (\ref{nlimit}) becomes 
\be\label{newton}
\frac{d^2 {\bf r}}{dt^2} = 
\left( - \frac{G_NM}{r^2} 
+ \frac{\Lambda}{3}r\right)\, \frac{\bf r}{r} \, .
\ee 
Eq. (\ref{newton}) is the non-relativistic acceleration 
acting on the test-particle and consists of the standard 
Newtonian term, which goes like $1/r^2$, plus a 
correction proportional to $r$, due to the cosmological 
constant $\Lambda$. This equation can also be written as
\be\label{newton2}
\frac{d^2 {\bf r}}{dt^2} =
- \frac{G_NM_{eff}}{r^2} \,\frac{\bf r}{r}\, ,
\ee
where $M_{eff}$ is the effective Newtonian mass enclosed 
within the radius $r$
\be\label{m-eff}
M_{eff}(r) = M - \frac{8}{3}\pi r^3 \rho_\Lambda
\ee
and $\rho_\Lambda$ is the energy density associated with
the cosmological constant: $\Lambda = 8\pi G_N\rho_\Lambda$. 
Since $M_{eff}$ depends on the distance from the source, 
for $\Lambda \neq 0$ the test-particle feels an effective 
violation of standard gravitational inverse square law.
As the matter of the fact, the new force is an inertial effect, 
related to the choice of the coordinate system. This 
becomes more evident in the limit $M=0$, where the 
test-particle continues to feel the force proportional
to its distance from the origin of the coordinate system, 
but where the latter is a point like all the others; the 
phenomenon disappears in a comoving reference frame.

\section{General features}
\label{s-discussion}

As we have seen in the previous section, in a static
coordinate reference frame, cosmological constant effects 
can be interpreted by a non-relativistic test-particle
as an effective violation of the gravitational inverse square 
law. However, one should not confuse this assertion with
present attempts of many authors focusing on deviations 
from standard gravity. 
There, deviations at small distances ($\lesssim 1$ mm) are 
usually referred to in order to solve the huge gap between the 
observed value of the Dark Energy and the one we could naively 
predict for the cosmological constant from particle physics 
considerations. On the other hand, deviations at larger scales 
($\gtrsim 1$ Gpc) would aim at explaining the present accelerated 
expansion rate of the universe without invoking Dark Energy, 
but just modifying General Relativity \footnote{There exist also 
attempts to modify gravity at the galactic scale, i.e. at distances
of order 10 -- 100 kpc, to explain galaxy rotation curves 
without Dark Matter.}. Here, the picture is more 
conservative: the framework is the one of the theory of
General Relativity with a small cosmological constant in 
the Newtonian limit and the target is to consider
Dark Energy effects on small distances.

Let us now discuss the main features emerging from this 
picture. From Eq. (\ref{newton2}) we can see that a
test-particle is attracted by a body of mass $M$ with
a weaker force (we take $\Lambda > 0$) than the case 
with null cosmological constant. Nevertheless,
for $\rho_\Lambda \approx 6 \cdot 10^{-30}$ g/cm$^3$
(the value we deduce from present cosmological data \cite{data}) 
the effect is so tiny that it is essentially impossible to
detect.

In order to make some estimates, it is convenient to
introduce the quantity
\be\label{beta}
\beta(r) = \frac{\frac{8}{3}\pi r^3 \rho_\Lambda}{M} \, ,
\ee
which is the ratio of the repulsive cosmological constant 
force to the attractive standard term. From Eq. (\ref{beta}), 
we can see that in laboratory experiments,
where for example $M \sim 100$ kg and $r \sim 100$ cm,
$\beta$ is about 10$^{-28}$. In the Solar System, 
with $M \sim M_\odot$ the Solar mass and 
$r \sim 10^{13}$ cm the mean Earth-Sun distance, $\beta$
is at the level of $10^{-23}$. The correction is so small that
relativistic effects may be much more important.

Of course, if Dark Energy was not the cosmological constant 
$\Lambda$, its value could vary from one point of the 
spacetime to another. In this case, we could put
phenomenological upper bound on its local magnitude,
even if probably this approach is not theoretically
well motivated and we should expect other more relevant
phenomena (for example violation of the universality of free
fall or spacetime variation of fundamental constants),
depending on the unknown origin of Dark Energy.

As the distance between test-particle and massive body increases,
$\Lambda$ repulsion term becomes more and more relevant
and $M_{eff}$ decreases. Gravitational attraction
dominates until when $M_{eff} > 0$. The distance $R$ for
which $M_{eff}(R) = 0$ is an unstable equilibrium point
and for $r > R$ the real spacetime expansion overcomes
the attractive gravitational force of the body of mass $M$.
For example, a test-particle feels an effective attraction
towards the Sun up to a distance $R \approx 100$ pc. As for
the Milky Way, whose mass is about 10$^{12}\;M_\odot$, 
the distance is $R \approx 1$ Mpc: this means that
the Local Group is a gravitationally bound system.
On the other hand, from this simple picture follows
that the Virgo Cluster is not exerting an effective
attractive force on us: its mass is $M \sim 10^{15}\;M_\odot$,
implying $R \sim 10$ Mpc, whereas it is at a distance
of about 20 Mpc from us. Here, however, the situation 
is more subtle, since the Virgo Cluster and the
Local Group are not two objects in an empty space,
but between them there are other galaxies and clusters,
whose effect may be to form a sort of ``chain'' or
``gravitationally bound filament'' (see the end of
Section \ref{s-conclusion}), so that they may be
part of the same bound system.

In this connection, we can work out the following simple
description for $N$ point-like massive particles which 
interact gravitationally. First, we choose a static
system of coordinates, whose origin appears as the source
of an effective radial repulsive force for all the particles.
Second, we consider the standard Newtonian gravitational 
force acting on each particle and equal to the sum
of all the gravitational forces exerted on the particle
by the other ones. The result is that the acceleration
of the $i$th particle is
\be
\frac{d^2{\bf r}_i}{dt^2} = -G_N \sum_{i \neq j} 
\frac{M_j}{r_{ij}^2}\frac{{\bf r}_{ij}}{r_{ij}} \,
+ \, \frac{\Lambda}{3} r_i \frac{{\bf r}_i}{r_i}\; ,
\ee
where $M_j$ is the (real) mass of the $j$th particle,
${\bf r}_{ij}={\bf r}_i - {\bf r}_j$ and $r_{ij}=|{\bf r}_{ij}|$.
This example shows clearly that the $\Lambda$ force
has to be an effective (or apparent) force, due to the choice
of the reference frame. In addition to this, considering
the limit $\sum \rar \int$, we can easily generalize 
Eq. (\ref{m-eff}) in the case of spherically symmetric 
mass distribution
\be
M_{eff}(r) = M(r) - \frac{8}{3}\pi r^3 \rho_\Lambda \, ,
\ee
where
\be
M(r) = \int_0^r \rho(x) \, 4\pi x^2\,dx \, ,
\ee
is the matter mass within the radius $r$ and $\rho(x)$
the matter mass density at the distance $x$ from the origin.

$N$-body simulations are often used to study cluster
structure. Particular interest is devoted to 
mass density profile and substructure, because it 
is believed that they retain informations
on the evolutionary history. They can also provide a
measurement of $\Omega_{mat} = \rho_{mat}/\rho_c$, the
matter energy density in the universe to the critical 
energy density ratio, but are (at least usually) 
essentially insensitive to $\Lambda$: for a typical 
cluster of mass $M \approx 10^{15}\;M_\odot$ and size 
1 Mpc, $\beta$ is no more than $10^{-3}$.

\section{Galaxy cluster masses}
\label{s-implications}

Galaxy clusters are the largest gravitationally bound
systems in the universe, containing usually some few
hundreds galaxies spread over a region of roughly
1 Mpc. At present there exist three independent 
methods to measure their masses, based respectively on
galaxy kinematics \cite{v-m}, X-ray profile \cite{x-m}
and gravitational lensing \cite{gl-m}.

The first approach focuses on galaxy motion within the 
cluster. Basically, we assume that the cluster is in
hydrostatic equilibrium and is spherically symmetric,
so that the acceleration of a galaxy (here considered as
a test-particle) at the distance $r$ from the cluster
center is
\be
\frac{v^2}{r} = \frac{G_NM(r)}{r^2} \, ,
\ee
where $v$ is the galaxy velocity and $M(r)$ the total
cluster mass within the radius $r$ (for more details, 
see e.g. Ref. \cite{v-m}).
For $\Lambda \neq 0$, the effective gravitational force 
is not provided by $M(r)$ but by $M_{eff}(r)$, so that
we really measure the latter quantity.

As for the second approach, the key point is that
the hot low-density intracluster gas is expected to
have a distribution similar to the one of the galaxies
in the cluster and to be able to trace the cluster 
gravitational potential of all the matter. Assuming 
that the gas is in hydrostatic equilibrium, we can 
write \cite{x-m2}
\be
\nabla P = - \rho \nabla \phi \, ,
\ee 
where $P$ is the gas pressure, $\rho$ the gas density
and $\phi$ the gravitational potential of the cluster.
If the latter is spherically symmetric
\be\label{pot}
\phi = - \frac{G_NM(r)}{r} \, .
\ee
However, for a non zero cosmological constant the Newtonian
gravitational potential is not exactly that given in 
Eq. (\ref{pot}), but we have to perform the substitution 
$M(r)\rar M_{eff}(r)$: even in this case we do not measure
the real cluster mass but also the $\Lambda$ contribution.

The last method is based on gravitational lensing and
can measure cluster masses from the produced distortion
of background galaxies. In this case the slow motion
approximation considered in this paper is clearly
inadequate and a relativistic treatment is necessary;
this can be found in the literature. The important
feature is that light deflection is not affected by a
non-zero cosmological constant \cite{ssystem, lensing},
implying that the method measures the ``real''
cluster mass $M$. 

Since the three independent techniques provide consistent 
cluster masses, typically within radii of about 1 Mpc, it
is common belief that we can reliably determine them with
an accuracy at the level of 30\%, the observed scatter
of the data. As for Dark Energy effects on these 
measurements, they are indeed usually negligible: for 
$\rho_\Lambda \approx 6 \cdot 10^{-30}$ g/cm$^3$, the 
theoretical ratio of galaxy kinematics or X-ray mass 
$M_{eff}$ to the gravitational lensing one $M$ 
within the radius $r$ is
\be\label{mm-ratio}
\frac{M_{eff}}{M} = 1 - 0.007 
\left(\frac{10^{14}\,M_\odot}{M}\right)
\left(\frac{r}{1\,{\rm Mpc}}\right)^3 \, .
\ee
Since standard value are $M \sim 10^{12}-10^{15}\;M_\odot$
and $r \sim 0.2-1$ Mpc, the discrepancy is irrelevant
for present accuracy.

However, if we are interested in the observation of
Dark Energy effects on these gravitationally bound systems,
we could select suitable galaxy clusters with features
favorable for our purpose. What we would need are
light and non-compact galaxy clusters: for example,
for $M \approx 10^{13}\;M_\odot$ and $r\approx 2$ Mpc
the mass measured by the first two methods with respect
to the gravitational lensing one should differ 
at the level of 60\%. Of course this
kind of measurements are challenging, but they are 
not impossible to reach.

An alternative approach is to measure cluster masses
through galaxy kinematics and to study the behavior
of the gravitational force \footnote{For this purpose,
X-ray profile is probably not competitive, since we
would need large volumes with very low matter density,
so that X-ray measurements are difficult.}: for $\Lambda \neq 0$
the gravitational force can not decrease faster than
$1/r^2$, whereas for $\Lambda > 0$ it can. Here we
would need a very compact cluster with few small 
satellites at larger distances which can be used to
determines $M_{eff}$ as a function of $r$.

There exist also the more favorable possibility 
that Dark Energy is not uniformity distributed
and that in some galaxy cluster $\rho_\Lambda$ is larger
than its mean value. For instance, it would be relatively
easy to observe Dark Energy effects if $\rho_\Lambda>10^{-27}$
g/cm$^3$, only 2 -- 3 order of magnitude larger than
its mean value. At the moment we can only say that
the general agreement between the three techniques rejects
a frequently intracluster cosmological constant of this
magnitude: even if from systems of size of about 1 Mpc,
it represents a constraint much stronger than the ones
coming from the Solar System \footnote{The statement
is strictly valid only if Dark Energy can be described
by a cosmological constant on scales around 1 Mpc, even
if not necessary with the value we deduce from cosmological
data. For observational signatures of other Dark Energy
candidates, see e.g. Ref. \cite{finelli}.}.

\section{Conclusion}
\label{s-conclusion}

Often it is assumed, without particular cure, that
cosmological constant effects enter into the dynamics of
the universe on large scales but that they are completely
negligible for the dynamics of gravitationally bound systems.
This is indeed true in general and in this paper I have
discussed in some detail the topic. Moreover, I have
shown that measurements of galaxy cluster masses can provide
evidences of Dark Energy in the dynamics of gravitationally
bound systems with typical size of 1 Mpc.
This kind of measurements would be very important for
a future solution of the Dark Energy puzzle and of the
mysterious accelerated expansion rate of the universe.
The key point is to find light and non-compact galaxy
clusters and then to be able to perform precise
mass measurements which are sensitive and insensitive
to $\Lambda$. An alternative possibility is to look for
very compact clusters with few distant satellites and
to measure the gravitational force behavior as a
function of the distance from the cluster center.

Here I have considered cosmological constant effects in
the Newtonian limit, i.e. under the assumptions of slow 
motion and weak gravitational field. First order 
relativistic corrections of Eq. (\ref{newton}) are 
suppressed by
\be\label{est-approx}
& \frac{G_NM}{r} & \sim 10^{-7} - 10^{-4} \, , \nonumber\\
& \frac{\Lambda}{3} \, r^2 & \sim 10^{-8} \, , \nonumber\\
& v^2 & \sim 10^{-6} \, ,
\ee
where the estimates are for $M \sim 10^{12} - 10^{15} \, M_\odot$,
$r \sim 0.2 - 1$ Mpc and $v \sim 300$ km/s. Corrections
of the same order of magnitude have to be expected in all 
the results coming from Eq. (\ref{newton}), so that 
the Newtonian limit is a good approximation for our 
purpose and there are no reasons to go beyond the
non-relativistic picture. If it was not so, exact General
Relativity equations should be used: in the special case 
of spherical symmetry, they reduce to ordinary differential 
equations (such as Eq. (20b) of Ref. \cite{frolov}) 
that can be solved numerically.

Finally, the Newtonian framework suggests us
the following simple picture of the universe.
Each galaxy can be thought of as a point-like particle
of mass $M$ at the center of a bubble of radius $R$, with
$M_{eff}(R)=0$. Inside the bubble, gravitational attraction
towards the galaxy overcomes the effective repulsive
force due to the cosmological constant $\Lambda$.
Gravitationally bound systems such as galaxy clusters are
essentially overlapping bubbles. On larger scale, the
universe can be seen as a space filled with larger
bubbles, representing galaxy clusters; if two bubbles
are not in direct contact, they exert an effective 
repulsive force each other. On the other hand, if
the distance between the center of two bubbles is
smaller than the radius of the larger bubble, the two
bubbles are surely exerting attractive force each 
other (even if at a given time their distance is increasing
because of the expansion of the universe, the 
acceleration of the relative separation is negative).
As the universe expands, 
$\Omega_\Lambda=\rho_\Lambda/\rho_c$ increases and 
islands of gravitationally bound systems will be more 
and more diluted. Structure formation goes on inside 
bubbles. If there exist ``chains'' of overlapping 
bubbles, they may be gravitationally bound and
collapse, generating super-bubbles, but disjointed
chains are destined to go away and not to interact.

\begin{acknowledgments}
I wish to thank Alexander Dolgov and Francesco Villante 
for useful comments and suggestions.
\end{acknowledgments}

\end{document}